\documentclass{optica-article}

\journal{opticajournal} 

\articletype{Research Article}

\usepackage{lineno}

\begin{document}

\title{Recirculating frequency-shifting loop for flexible optical chirp generation and FMCW LiDAR}

\author{Alexander Mrokon,\authormark{1,*} Sebastian Schöler,\authormark{1} Leonard Vossgrag,\authormark{1} Karsten Buse,\authormark{1,2} and Ingo Breunig\authormark{1,2}}

\address{\authormark{1}Laboratory for Optical Systems, Department of Microsystems Engineering - IMTEK, University of Freiburg, Georges-Köhler-Allee 102, Freiburg, 79110, Germany\\
\authormark{2}Fraunhofer Institute for Physical Measurement Techniques IPM, Georges-Köhler-Allee 301, Freiburg, 79110,Germany\\}

\email{\authormark{*}alexander.mrokon@imtek.uni-freiburg.de} 


\begin{abstract*} 
A recirculating frequency-shifting loop (FSL) provides a highly flexible platform for generating coherent optically chirped light with tunable bandwidth, duration, chirp rate and repetition rate. The properties of the chirped light are controlled using low-frequency sinusoidal electronic drive signals, enabling deterministic waveform synthesis without complex high-speed electronics. We achieve chirp bandwidths of 10~GHz with a duration in the nanosecond regime, representing one of the fastest tunable laser sources to date. Using such chirped laser pulses, we demonstrate coherent FMCW LiDAR measurements over distances up to 3~m, highlighting the potential of FSL-based sources for compact, scalable and high-performance ranging systems.
\end{abstract*}

\section{Introduction}

Continuous, mode-hop-free tuning of laser light is essential for a wide range of scientific and technological applications. High-resolution laser spectroscopy plays a critical role in health and environmental monitoring through the detection and analysis of molecules \cite{PanAdvances}. Accurate distance and velocity measurements are achieved using frequency-modulated continuous-wave (FMCW) LiDAR systems \cite{BoserLidar}, and swept-source optical coherence tomography (OCT) is widely used in medical imaging \cite{FujimotoTheEcosystem}. 
Recent work has demonstrated that positronium can be cooled to ultralow velocities (~1 K) within its short lifetime using a chirped laser pulse train that dynamically tracks the Doppler shift \cite{shuCoolingPos}.
Consequently, linear tuning of laser light is of fundamental importance not only for basic scientific research but also for a broad spectrum of practical, real-world applications.

Current-modulated DFB lasers offer a simple and fast tuning mechanism over several tens of gigahertz, but suffer from intrinsic sweep nonlinearity\cite{Li20, NehmetallahLarge-Volume}. 
MEMS-tuned VCSELs enable ultrafast frequency sweeps spanning multiple tens of terahertz, although additional linearization techniques are required and the coherence length is typically limited to the meter scale \cite{MingHigh-Accuracy,JayaramanVCSEL,MingLaserFreq,IiyamaThreee-Dimensional,Jayaraman14, Khan21}. 
Akinetic semiconductor swept sources provide fully electronic and linear tuning over comparable bandwidths with coherence lengths below 1~m, making them well suited for optical coherence tomography \cite{DrexlerAkinetic,LeitgebAkinetic}. 

In contrast, resonator-enhanced electro-optic tuning via self-injection locking combines highly linear, ultrafast sweeps with coherence lengths extending to hundreds of meters, but the achievable tuning rate is fundamentally constrained by the resonators quality factor \cite{KippenbergUltrafast,KippenbergHighDensity,VahalaPockelsLaser,Kondratiev17}. 

An alternative, laser-medium-independent strategy for frequency conversion is adiabatic frequency conversion (AFC), in which light stored inside an optical resonator is frequency-shifted by modulating the cavity eigenfrequency on a timescale smaller than the photon decay time \cite{SatoshiWavelength}. While this approach enables ultrafast frequency shifts without direct modification of the laser source, it is fundamentally constrained by the finite photon lifetime within the resonator \cite{LipsonChanging,YannickPockels,CardenasAdiabatic,MrokonAFCFMCW}. In addition, the tuning process can be affected by unwanted nonlinearities arising from mechanical resonances of the resonator material \cite{MrokonPiezo}.

Several approaches to continuous laser frequency tuning have been developed, each involving inherent trade-offs among tuning range, tuning speed, linearity, and coherence length. Despite their different physical implementations, these techniques share a common characteristic: they rely on complex and often application-specific electronic control systems to realize and stabilize the tuning mechanism. 

Such electronic is typically sophisticated and implemented using an arbitrary waveform generator (AWG) that provides the linear voltage or current drive required for the generation of linearly-chirped laser light.
This increases the overall system complexity and introduces costly high-performance electronic components.
Thus, scalability and robustness are reduced. Furthermore, the reliance on external electronic drivers complicates photonic integration and limits suitability for compact or cost-sensitive implementations.

Approaches that decouple chirp generation from the direct tuning of the laser and enable wideband operation using comparatively simple electronics are essential to facilitate a large-scale deployment of applications such as high-resolution FMCW LiDAR, precision metrology, and ultra-fast spectroscopy. 
In this context, recirculating frequency-shifting loop (FSL) architectures driven by a single continuous-wave laser have recently demonstrated the generation of broadband chirped waveforms using a single sinusoidal low-frequency electronic control signal \cite{ChatellusReconfigurable}. 
Their applicability in spectroscopy \cite{DuranSpectroscopy}, photonic computing \cite{WuUltraLowLatencySpectralResolving, ChatellusOpticalReal-timeFourier}, and in the generation of arbitrary radio-frequency (RF) waveforms \cite{ChatellusRFAWG} has been demonstrated in previous work.
However, a detailed investigation of the influence of the modulation frequency on the chirp configuration, including a rigorous analysis of the generated linear chirps for distance measurement applications, has not yet been reported. 

We demonstrate that photonic FSLs provide a viable pathway toward scalable, coherent, and broadband chirp generation without the need for high-speed electronic control or laser-specific tuning mechanisms. 
An FSL based on an acousto-optic modulator (AOM) enables flexible control of chirp bandwidth and tuning time, making it well suited for FMCW LiDAR systems. 
Moreover, employing in-phase/quadrature (I/Q) modulators operated in single-sideband (SSB) mode provides enhanced control of the chirp rate and direction, allowing precise waveform tailoring without increasing electronic complexity.

\section{Operating principle of the frequency-shifting loop}

\begin{figure}[htpb]
\centering
\includegraphics[width=\linewidth]{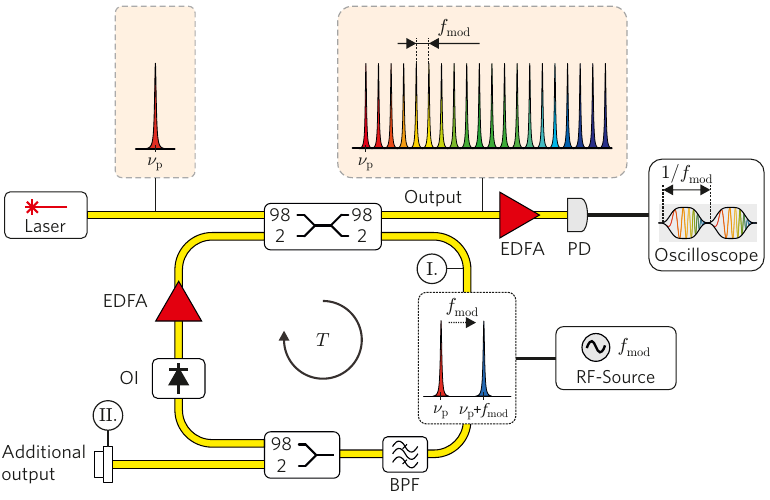}
\caption{Schematic of the frequency-shifting loop (FSL). Light at frequency $\nu_{\mathrm{p}}$ from a continuous-wave seed laser is injected into a fiber loop with round-trip time $T$, where the circulating field undergoes a controlled frequency shift of $ f_{\mathrm{mod}}$ on each round trip while an EDFA compensates for loop losses. A bandpass filter (BPF) and an optical isolator (OI) ensure stable operation. After multiple circulations, the loop generates a discrete optical spectrum corresponding to linearly chirped pulses, which can be detected using a photodiode (PD) and oscilloscope. An additional output provides direct access to the frequency-chirped optical signal without the pump light.}
\label{figSetup}
\end{figure}

A frequency-shifting loop (FSL) enables the generation of a frequency comb through repeated circulation of laser light in an optical loop \cite{ChatellusTheory, ChatellusGeneration, YatsenkoFSFlaser}. 
As illustrated in Fig.~\ref{figSetup}, light from a continuous-wave laser is injected into a fiber loop via an optical coupler. 
The round-trip time of the loop is denoted by $T$.
During each round trip, the optical field experiences a frequency shift $f_{\mathrm{mod}}$ imposed by a frequency-shifting element, while optical amplification compensates for loop losses.
After $n$ circulations, the loop output exhibits discrete spectral components in the frequency domain, indexed by $n$ and spaced by $f_{\mathrm{mod}}$, with a total bandwidth determined by the effective number of loop round trips.
In practice, the achievable comb bandwidth is limited by the optical components in the loop. In particular, an optical bandpass filter (BPF) constrains the spectral extent of the circulating field and suppresses amplified spontaneous emission (ASE) noise originating from the erbium-doped fiber amplifier (EDFA), thereby ensuring stable operation. In addition, an optical isolator (OI) is used to prevent undesired back reflections within the loop.
For the analysis of the time signal, the generated pulses can be detected by combining them with a reference derived from the original seed laser in the optical coupler. This  process converts the optical frequency into a radio-frequency (RF) waveform, which is detected using a photodiode and an oscilloscope. Furthermore, a fraction of the light in the loop is extracted via an additional output coupler (Output~II in Fig. \ref{figSetup}), providing direct access to the optical signal in the loop without interference from the seed laser. 
It should be noted that the entire loop is implemented using polarization-maintaining fibers.

The mode-locking frequency for a bandwidth-limited optical pulse is given by $m/T$. Here, $m$ is a positive integer denoting the number of pulses circulating in the loop.
When successive frequency shifts are applied in the loop with a slight detuning from the mode-locking condition, i.e., $f_{\mathrm{mod}} \neq m/T$, the system operates in a regime where optical chirps can be generated. This is described in detail by Guillet de Chatellus et al. in \cite{ChatellusReconfigurable}.
The effective frequency detuning from the mode-locked frequency comb is then given by
\begin{equation}
\Delta f_{\mathrm{mod}} = f_{\mathrm{mod}} - \frac{m}{T}.
\end{equation}
As a result, the output optical field no longer consists of bandwidth-limited pulses, but instead forms a periodic sequence of temporally stretched optical pulses with a repetition rate of $f_{\mathrm{mod}}$.
Accordingly, the chirp bandwidth is given by 
\begin{equation}
\text{chirp bandwidth}(f_{\mathrm{mod}})
= N f_{\mathrm{mod}} .
\end{equation}
The integer $N$ thus represents the number of frequency components that can be sustained in the loop \cite{ChatellusReconfigurable}.

\begin{figure}[htpb]
\centering
\includegraphics[width=\linewidth]{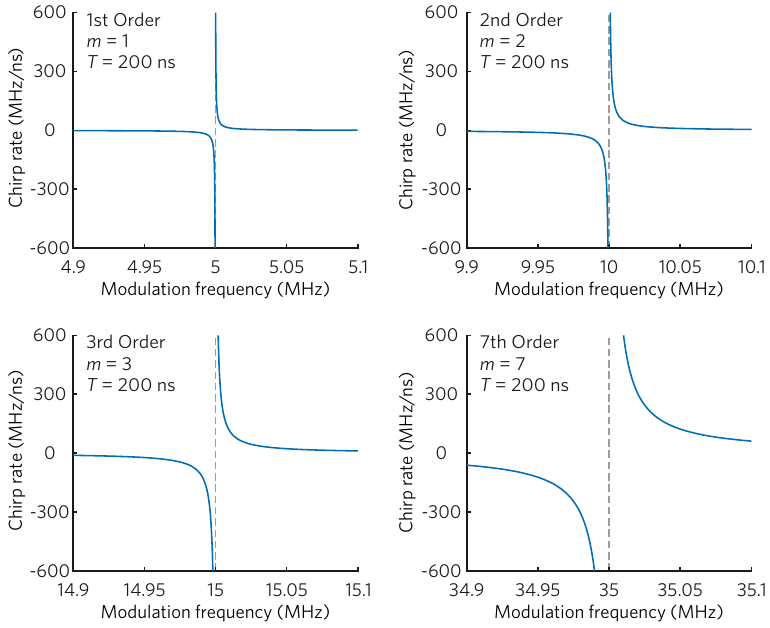}
\caption{Simulated chirp rate versus modulation frequency for a round-trip time \(T = 200~\mathrm{ns}\) and orders \(m = 1, 2, 3,\) and \(7\). The dashed lines indicate the mode-locking frequencies \(f_{\mathrm{mod}} = m/T\). The chirp rate diverges near the mode-locking condition and changes sign across it.
This behavior illustrates the strong dependence of the achievable chirp rate on both the modulation frequency \(f_{\mathrm{mod}}\) and the order \(m\).}
\label{figSim}
\end{figure}

Furthermore, the modulation frequency not only determines the chirp bandwidth but also sets the order $m$, which specifies the harmonic of the fundamental loop frequency \(1/T\) at which the system operates.
Higher orders correspond to larger modulation frequencies and, consequently, to an increased chirp bandwidth.

However, higher orders lead to a reduction in the maximum chirp duration, as \(m\) pulses share the round-trip time, resulting in a maximum pulse duration of \(T/m\).

This behavior follows directly from the expression
\begin{equation}
\text{chirp duration}(f_{\mathrm{mod}})
= N  \frac{\Delta f_{\mathrm{mod}}}{f_{\mathrm{mod}}}  T
= N  \frac{ f_{\mathrm{mod}}  - \frac{m}{T}}{f_{\mathrm{mod}}}  T .
\end{equation}
The chirp duration is therefore proportional to the relative detuning \(\Delta f_{\mathrm{mod}} / f_{\mathrm{mod}}\).
For higher-order operation, the mode-locking frequency \(m/T\) increases with \(m\), such that the corresponding modulation frequency \(f_{\mathrm{mod}}\) also becomes larger. Even if the absolute detuning \(\Delta f_{\mathrm{mod}}\) is kept constant, the ratio \(\Delta f_{\mathrm{mod}} / f_{\mathrm{mod}}\) decreases with increasing modulation frequency. As a result, the chirp duration becomes shorter for higher orders.

From these considerations, it follows that the chirp rate generated in an FSL depends strongly on the applied modulation frequency \(f_{\mathrm{mod}}\) and its detuning \(\Delta f_{\mathrm{mod}}\) from the mode-locking frequency \(m/T\).
The magnitude and sign of this detuning determine both the slope and the direction of the generated chirp.
Following the description in~\cite{ChatellusReconfigurable}, the chirp rate as a function of the modulation frequency \(f_{\mathrm{mod}}\) can be expressed as
\begin{equation}
\text{chirp rate}(f_{\mathrm{mod}})=\frac{\mathrm{bandwidth}}{\mathrm{duration}}=\frac{N  f_{\mathrm{mod}}}{ N \frac{\Delta f_{\mathrm{mod}}}{f_{\mathrm{mod}}}  T}=\frac{f_{\mathrm{mod}}^2}{T \Delta f_{\mathrm{mod}}}
=\frac{f_{\mathrm{mod}}^2}{T\left(f_{\mathrm{mod}}-\frac{m}{T}\right)}.
\end{equation}

This illustrates a fundamental trade-off of the FSL architecture. Operating at higher order \(m\) increases the achievable chirp bandwidth, since the chirp bandwidth scales strongly with \(f_{\mathrm{mod}}\). At the same time, however, the temporal window over which the chirp is generated becomes smaller. 
Furthermore, the repetition rate of the generated chirped waveform increases with the order, as it is directly linked to the modulation frequency. 
As a result, higher-order operation yields chirps with increased bandwidth, reduced duration, and higher repetition rates.

To illustrate the influence of \(f_{\mathrm{mod}}\) and \(m\), the chirp rates were displayed in Figure \ref{figSim} for an effective frequency detuning of \(\Delta f = -100\ldots 100~\mathrm{kHz}\).
An example set of design parameters is obtained by considering a round-trip time of \(T = 200~\mathrm{ns}\). Under the mode-locking condition, the modulation frequency is determined by the integer order \(m\) according to \(f_{\mathrm{mod}} = m/T\). 
For order numbers \(m = 1\), \(m = 2\), \(m = 3\), and \(m = 7\), this results in modulation frequencies of \(f_{\mathrm{mod}} = 5~\mathrm{MHz}\), \(10~\mathrm{MHz}\), \(15~\mathrm{MHz}\), and \(35~\mathrm{MHz}\), respectively.

For a fixed detuning $\Delta f_{\mathrm{mod}}$ of 10~kHz from the modulation frequency for mode-locked pulses, the resulting chirp rates increase significantly with order, yielding $12.5~\mathrm{MHz/ns}$, $50~\mathrm{MHz/ns}$, $112.5~\mathrm{MHz/ns}$, and $612.5~\mathrm{MHz/ns}$, respectively. 
It is worth noting that such a frequency offset is not technically demanding. 
In practice, a detuning of the order of $10~\mathrm{kHz}$ can be easily implemented using simple and widely available electronic components, such as standard function generators or low-cost frequency control circuitry from radio broadcast systems.
This makes the chosen detuning both experimentally accessible and robust. 
Assuming $N = 100$ spectral components, the corresponding chirp durations are $39.9~\mathrm{ns}$, $19.98~\mathrm{ns}$, $13.32~\mathrm{ns}$, and $5.71~\mathrm{ns}$, while the achieved bandwidths are $501~\mathrm{MHz}$, $1.001~\mathrm{GHz}$, $1.501~\mathrm{GHz}$, and $3.501~\mathrm{GHz}$ for $m = 1$, $2$, $3$, and $7$, respectively. 
These results highlight that the chirp rate can be tuned over orders of magnitude using simple electronic control.

In an FMCW LiDAR system, the chirp bandwidth determines the distance resolution, while the chirp duration sets the maximum measurable range. Higher orders increase the repetition rate, enabling faster acquisition. All these parameters can be controlled via a single parameter, the  modulation frequency \(f_{\mathrm{mod}}\). In the following, we use this approach to provide a proof of principle of the linearity and achievable optical output power of the generated chirps.

In conclusion, the chirp characteristics are fully determined by the modulation frequency $f_{\mathrm{mod}}$, which sets both the detuning from the mode-locking condition and the order $m$, thereby enabling precise control of bandwidth, duration, and repetition rate.
This enables the generation of wideband, linear, and coherent frequency sweeps under low-frequency electronic control.

\section{AOM-based frequency-shifting loop}

\begin{figure}[htpb]
\centering
\includegraphics[width=\linewidth]{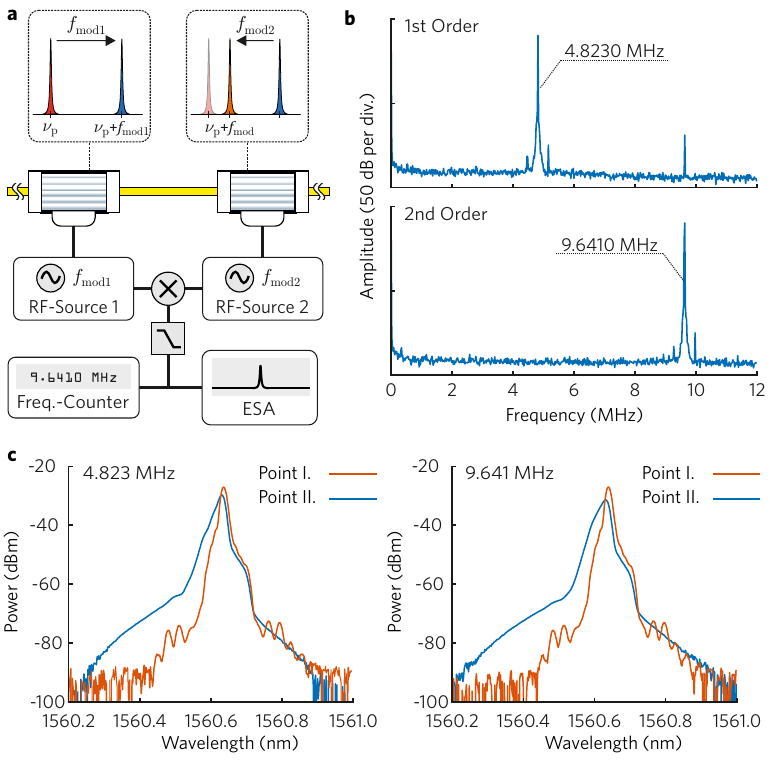}
\caption{ \textbf{a} Experimental schematic of the frequency-shifting loop (FSL) using two acousto-optic modulators (AOMs) driven by independent RF sources. The effective modulation frequency \(f_{\mathrm{mod}}\) is monitored with an electrical mixer, a low-pass filter, a frequency counter, and an electrical spectrum analyzer (ESA). \textbf{b} Electrical spectra for first- and second-order operation, showing peaks at \(f_{\mathrm{mod}}=4.8230~\mathrm{MHz}\) and \(f_{\mathrm{mod}}=9.6410~\mathrm{MHz}\). \textbf{c} Optical spectra measured at two extraction points of the loop (Point~I. and Point~II.), exhibiting spectral broadening due to the accumulated frequency shifts.
}
\label{figAOMsetup}
\end{figure}

To experimentally demonstrate the flexibility of this concept, we implemented an FSL employing two acousto-optic modulators (AOMs) as the frequency-shifting elements.The experimental setup is shown schematically in Fig.~\ref{figAOMsetup}a.
A narrow-linewidth continuous-wave laser was injected into the loop, which had a round-trip time of \(T = 207.53~\text{ns}\).
The first AOM imposed a fixed frequency shift of \(f_{\mathrm{mod1}} = 129.823~\mathrm{MHz}\) for first-order operation and \(f_{\mathrm{mod1}} = 134.641~\mathrm{MHz}\) for second-order operation, respectively, while the second AOM imposed a fixed shift of \(f_{\mathrm{mod2}} = 125~\mathrm{MHz}\). 
These values represent the maximum achievable orders with the AOM-based implementation, as the tunability of each AOM is limited to approximately 10~MHz. This enables the realization of first- ($m = 1$) and second-order ($m = 2$) modulation frequencies of the loop.
The net frequency shift accumulated per round trip,
\begin{equation}
f_{\mathrm{mod}}=\vert -f_{\mathrm{mod1}}+f_{\mathrm{mod2}}\vert=0\ldots 10~\mathrm{MHz},
\end{equation}
therefore amounted to \(f_{\mathrm{mod}}=4.823~\mathrm{MHz}\) and \(f_{\mathrm{mod}}=9.641~\mathrm{MHz}\), respectively. 

As two independent RF sources were required to drive the AOMs, we employed an electrical mixer followed by a low-pass filter to generate the difference frequency, which was monitored using a frequency counter and an electrical spectrum analyzer (ESA).

Figure~\ref{figAOMsetup}b shows the electrical spectra measured by the ESA and the frequency counter. For \(f_{\mathrm{mod}}=4.823~\mathrm{MHz}\), a clear first-order beat note at \(f_{\mathrm{mod}}\) is observed. A second-order component appears at \(2f_{\mathrm{mod}}=9.641~\mathrm{MHz}\), consistent with the expected accumulation of the frequency shift in the loop.

The corresponding optical spectra are shown in Fig.~\ref{figAOMsetup}c for both modulation frequencies. The spectra were recorded at two different extraction points of the loop (Point~I and Point~II), as indicated in Fig.~\ref{figSetup}. In both cases, a clear spectral broadening is observed, reflecting the cumulative frequency shifts induced by successive round trips in the loop.

\begin{figure}[htpb]
\centering
\includegraphics[width=\linewidth]{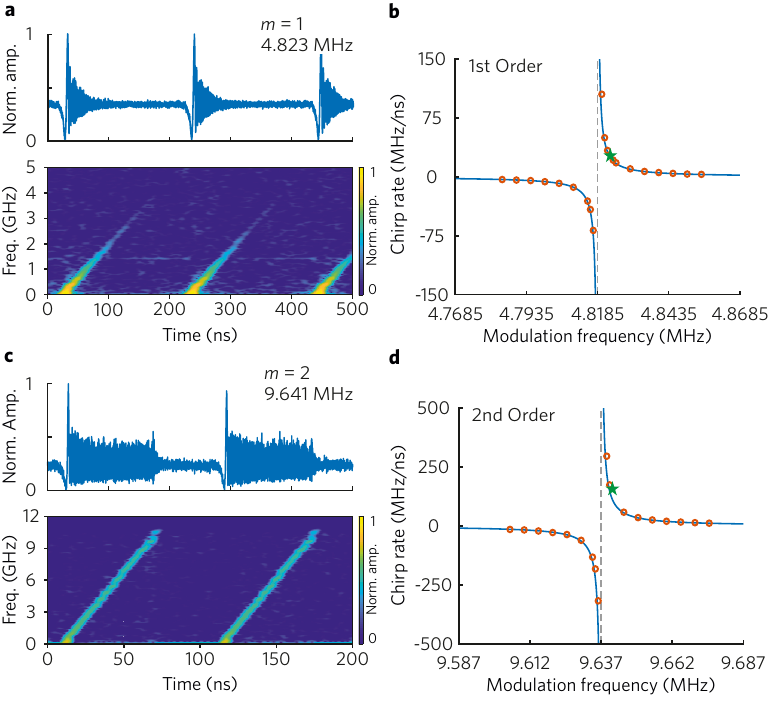}
\caption{\textbf{a} Normalized beat signals and corresponding spectrograms for first- and second-order operation at modulation frequencies of 4.823~MHz and 9.641~MHz, respectively. \textbf{b} Chirp rates as a function of modulation frequency for first- and second-order components, exhibiting a sharp change in magnitude and sign near the mode-locking condition. The chirp rates corresponding to 4.823~MHz and 9.641~MHz are indicated by green stars.
}
\label{figAOMChirpRate}
\end{figure}

For first-order operation, approximately -27.1~dBm of optical power is coupled into the loop via the coupler. The light extracted at Point~II exhibits a maximum power of approximately -29.7~dBm. This reduction in optical power is mainly attributed to the insertion loss of the two AOMs, which is approximately 5.7~dB.
For second-order operation, a comparable optical power of approximately -27.1~dBm is coupled into the loop. The light extracted at Point~II then exhibits a reduced maximum power of approximately -31.4~dBm. The increased power reduction is primarily attributed to the slightly higher combined insertion loss of the two AOMs, amounting to approximately 6.3~dB.

Individual spectral lines are not resolved due to the limited resolution bandwidth of the optical spectrum analyzer (OSA), which is approximately 5~GHz. Consequently, the measured spectra correspond to the envelope of the underlying discrete frequency components.

The observed spectral envelopes are primarily governed by the round-trip losses in the loop, which limit the number of frequency-shifted replicas that contribute significantly to the spectrum. In contrast, the optical bandpass filter, with a bandwidth of 10~GHz, suppresses amplified spontaneous emission (ASE) noise and thereby limits its impact on the measured spectrum.


Figure~\ref{figAOMChirpRate}a shows the temporal signal and the corresponding spectrogram for first-order operation at a modulation frequency of \(f_{\mathrm{mod}} = 4.823~\mathrm{MHz}\). The upper panel displays the normalized temporal waveform, where periodic pulses corresponding to successive round trips in the loop are observed. The spectrogram reveals a linear frequency sweep within each pulse interval. For first-order operation, the instantaneous frequency increases linearly by approximately 3~GHz, corresponding to the chirp bandwidth \(\Delta \nu\), over a chirp duration \(\Delta t\) of 103~ns. This results in a chirp rate of approximately 29~MHz/ns.

The dependence of the chirp rate on the modulation frequency for first-order operation is shown in Fig.~\ref{figAOMChirpRate}b. The extracted chirp rates exhibit a clear dependence on the modulation frequency and change sign close to the mode-locking condition. The operating point corresponding to the measurement shown in Fig.~\ref{figAOMChirpRate}a is indicated by a green star.

Second-order operation is shown in Fig.~\ref{figAOMChirpRate}c for a modulation frequency of \(f_{\mathrm{mod}} = 9.641~\mathrm{MHz}\). Similar to the first-order case, periodic pulses are observed in the temporal waveform, and the spectrogram shows a linear frequency sweep within each pulse interval. In this case, the frequency sweep extends over approximately 10~GHz within 63~ns, resulting in a significantly higher chirp rate of 158~MHz/ns.

The corresponding chirp-rate dependence on the modulation frequency for second-order operation is presented in Fig.~\ref{figAOMChirpRate}d. As in the first-order case, the chirp rate varies strongly with the modulation frequency and changes sign near the mode-locking condition. The operating point used for the spectrogram measurement in Fig.~\ref{figAOMChirpRate}c is again indicated by a green star.

The experimental scaling behavior of both orders is in good agreement with the theoretical estimation. 
As predicted by the expression of the chirp rate, the magnitude of the chirp rate increases rapidly as the modulation frequency approaches the mode-lock condition \(f_{\mathrm{mod}} \approx m/T\), where the detuning \(\Delta f_{\mathrm{mod}}\) becomes small. This trend is observed for both first- and second-order components, with the second-order data exhibiting consistently larger chirp-rate magnitudes over the same modulation-frequency range. The agreement between the measured chirp rates and the analytical scaling confirms that the frequency evolution in the FSL is governed by the accumulated frequency shift per round trip.

These results highlight the flexibility of the AOM-based FSL for generating tunable chirped waveforms using low-frequency electronic control. 
However, the experimental setup is subject to two main limitations. First, the AOM drivers allowed the output frequency around 130~MHz to be adjusted only in discrete steps of 1~kHz, thereby limiting how precisely the modulation frequency could be tuned to the exact mode-locking condition. 
Second, due to the limited tuning range of the AOMs of only 10~MHz, operation in this implementation is restricted to the first- and second-order regimes.

\section{Frequency-shifting loop with I/Q-modulator-based SSB generation}

\begin{figure}[htpb]
\centering
\includegraphics[width=\linewidth]{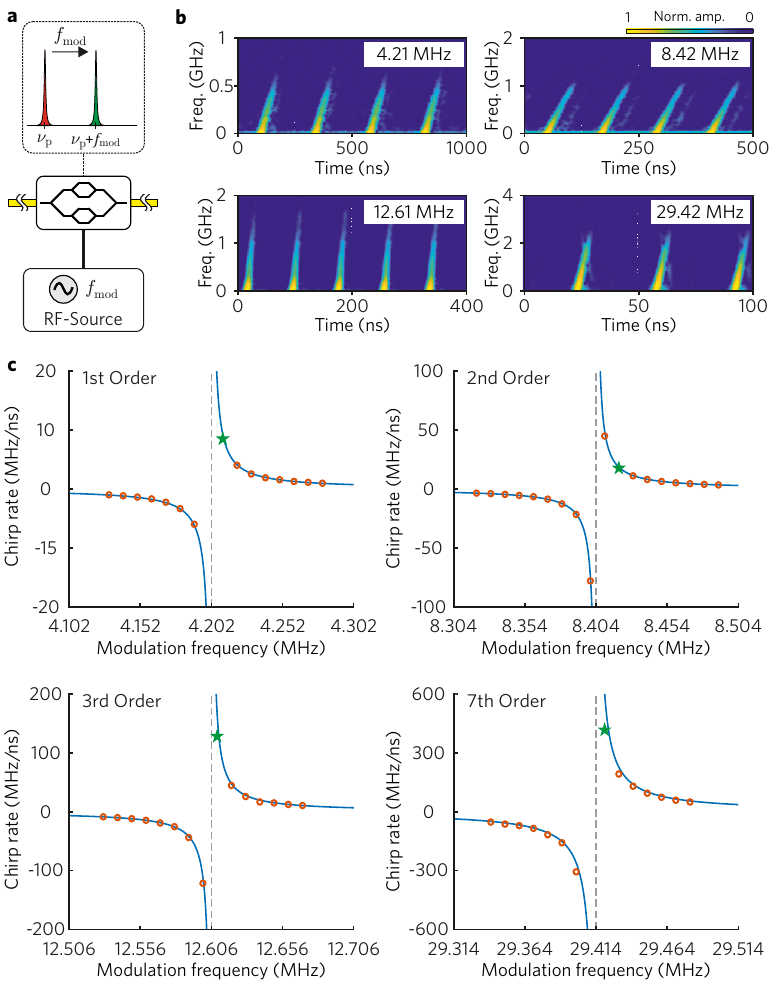}
\caption{\textbf{a} Schematic of the FSL implementation using an electro-optic I/Q modulator operated in single-sideband (SSB) mode. \textbf{b} Spectrograms of the loop output for first-, second-, third-, and seventh-order operation at modulation frequencies of 4.21~MHz, 8.42~MHz, 12.61~MHz, and 29.42~MHz, respectively, showing linear frequency sweeps. \textbf{c} Chirp rates as a function of modulation frequency for the corresponding orders, with green stars indicating the spectrograms.}
\label{figSSB}
\end{figure}

To enable the investigation of higher-order components beyond the first and second order, a second set of experiments was performed in which the acousto-optic modulators were replaced by an electro-optic I/Q modulator operated in single-sideband (SSB) mode, as illustrated in Fig.~\ref{figSSB}a. This approach overcomes the limited modulation bandwidth of the AOMs, which is approximately 10~MHz in our implementation, and allows access to higher-order frequency components within the FSL. 
By generating a pure optical SSB signal, the I/Q modulator provides a well-defined frequency shift per round trip of the loop.

As a result of the modified experimental configuration, the round-trip time of the loop changes to \(T = 237.98~\text{ns}\).

Figure~\ref{figSSB}b presents spectrograms of the detected loop output for modulation frequencies corresponding to the first, second, third, and seventh order. Specifically, the modulation frequencies are \(f_{\mathrm{mod}} = 4.21~\mathrm{MHz}\) for \(m = 1\), \(8.42~\mathrm{MHz}\) for \(m = 2\), \(12.61~\mathrm{MHz}\) for \(m = 3\), and \(29.42~\mathrm{MHz}\) for \(m = 7\).

The measured spectrograms for the different modulation frequencies are shown in Fig.~\ref{figSSB}b. 
For all investigated modulation frequencies, a sequence of repeated, linearly chirped frequency sweeps is observed.

As the modulation frequency is increased from 4.21~MHz to 29.42~MHz, the frequency bandwidth within each sweep increases, while the sweep duration correspondingly decreases. This behavior reflects the increasing frequency bandwidth determined by both the number of effective round trips and the applied modulation frequency, and consequently the reduced temporal period associated with higher modulation frequencies.

The linearity of the frequency evolution within each sweep confirms the deterministic and well-controlled frequency shifting provided by the single-sideband modulation scheme, enabling access to higher-order frequency components in the FSL. 
However, it should also be noted that the achieved frequency bandwidth is approximately eight to ten times smaller than that obtained with the AOM-based FSL. 
This reduction is attributed to the significantly higher losses of the I/Q modulator, which amount to approximately 16 dB and limit the number of effective round trips contributing to the frequency sweep.

The dependence of the chirp rate on the modulation frequency for the first, second, third, and seventh order is summarized in Fig.~\ref{figSSB}c. 
For each order, the chirp rate can be varied strongly with the modulation frequency and exhibits a characteristic divergence near the mode-lock condition, where the sign of the chirp rate changes.
With increasing order, the achievable chirp rates increase significantly, extending from a few tens of megahertz per nanosecond for the first order to several hundred megahertz per nanosecond for the seventh order, consistent with the cumulative nature of frequency shifting in the loop.
The spectrograms shown in Fig.~\ref{figSSB}b are marked by green stars.

These results demonstrate that I/Q-modulator-based single-sideband frequency shifting substantially extends the accessible modulation bandwidth and enables the controlled generation of higher-order chirped waveforms. In contrast to acousto-optic modulators, electro-optic modulators can readily support modulation frequencies well above 100~GHz. Moreover, implementation on thin-film lithium niobate platforms allows such operation to be achieved at CMOS-compatible drive voltages, making this approach attractive for scalable and integrated FSL realizations~\cite{LoncarCMOS}.

This approach would enable nanosecond-scale chirps with bandwidths in the multi-tens-of-terahertz range, using comparatively simple electronics and without the need for electrical amplifiers.
Thus, the electro-optic–modulator-based approach provides a versatile and scalable alternative to AOM-based implementations, particularly for applications requiring ultra-fast chirps with large bandwidths.

\section{FSL-based coherent ranging}

In the final part of the experiment, the linearity of the generated chirped waveforms was evaluated by integrating the frequency-shifting loop into a FMCW LiDAR system. By using the additional FSL output II. as the chirped optical source, the frequency-chirped light could be directly assessed under coherent ranging conditions.

Due to the reduced usable chirp bandwidth of the I/Q-modulator-based SSB implementation, the AOM-based FSL was employed as the chirp source for the FMCW LiDAR demonstration. Nevertheless, FMCW LiDAR measurements up to 3~m were achieved using electro-optically generated chirps. However, the resulting distance resolution was approximately four times lower than that of the AOM-based implementation, owing to the limited maximum chirp bandwidth of 2.5~GHz.
Therefore, to illustrate the versatility of the FSL concept for ranging applications, distance measurements were performed using both first- and second-order operation in the AOM-based setup.
Specifically, the modulation frequencies \(f_{\mathrm{mod}} = 4.823~\mathrm{MHz}\) and \(f_{\mathrm{mod}} = 9.641~\mathrm{MHz}\), as shown in Fig.~\ref{figAOMChirpRate}a and c, were used for the first- and second-order measurements, respectively.

\begin{figure}[htpb]
\centering
\includegraphics[width=\linewidth]{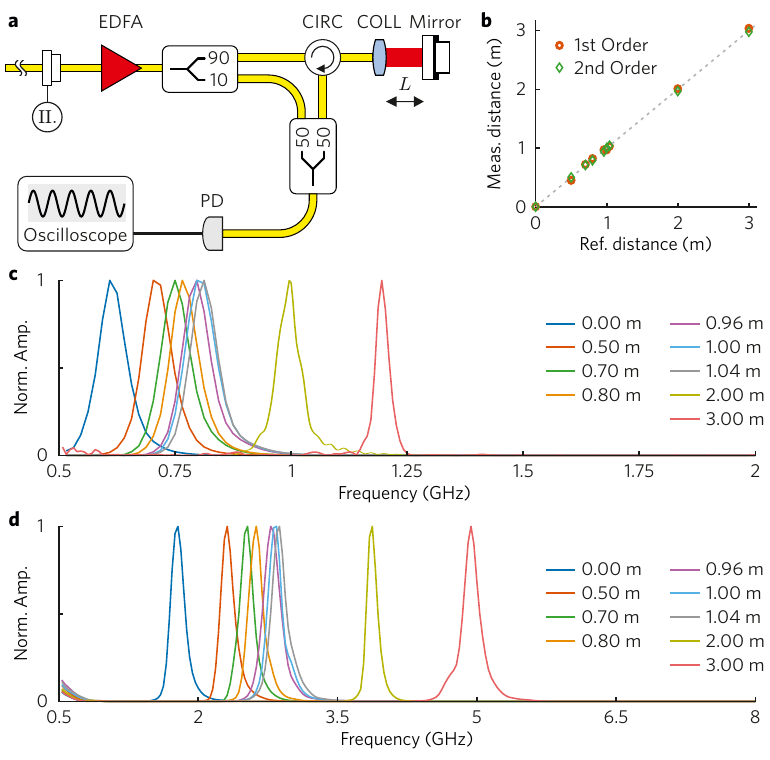}
\caption{\textbf{a} Experimental FMCW LiDAR setup using the AOM-based FSL as the chirp source. \textbf{b} Measured versus reference distances for first- and second-order operation, showing good linear agreement. \textbf{c} FFT spectra of the beat signals for different target distances in first-order operation. \textbf{d} Corresponding beat spectra for second-order operation, exhibiting higher beat frequencies due to the increased chirp bandwidth.}
\label{figFMCW}
\end{figure}

The experimental setup used for the FMCW LiDAR demonstration is shown schematically in Fig.~\ref{figFMCW}a. 
The chirped optical signal generated by the FSL is first amplified to 10~mW using an EDFA and then split by a 90:10 coupler. 
The majority of the optical power is directed through an optical circulator toward a collimator and a mirror, which serves as the ranging target. 
The light reflected from the mirror is routed back through the circulator and combined with a reference signal at a 50:50 coupler. The resulting interference signal is detected by a photodiode and recorded using a real-time oscilloscope (Keysight UXR0104B) for analysis. 

We used the real-time oscilloscope to perform the fast Fourier transform (FFT) of the detected beat signal. From the resulting spectra, the beat frequencies \(f_{\mathrm{B}}\) were extracted, as shown in Fig.~\ref{figFMCW}c and d.

For first-order operation, the extracted beat frequencies range from 0.61~GHz to 1.2~GHz, with measured values of 0.61, 0.70, 0.75, 0.77, 0.80, 0.80, 0.81, 1.0, and 1.2~GHz.

For second-order operation, the corresponding beat frequencies span a higher range, from 1.78~GHz up to 4.94~GHz, with measured values of 1.78, 2.31, 2.53, 2.62, 2.78, 2.84, 2.88, 3.87, and 4.94~GHz.

As the target distance increases, the beat frequency shifts proportionally, while the spectral linewidths remain narrow.
The increased beat frequencies observed for the second-order case are consistent with the higher effective chirp rate.

Using these frequency values together with the chirp bandwidth \(\Delta \nu\) the chirp duration \(\Delta t\) and the speed of light \(c\), the distance \(L\) between the collimator and the mirror can be calculated according to
\begin{equation}
L = \frac{c}{2} \left( \frac{\Delta t}{\Delta \nu} \left[ f_{\mathrm{B}} - f_{\mathrm{B},0} \right] \right),
\end{equation}
where \(f_{\mathrm{B},0}\) denotes the offset frequency at the collimator, arising from the difference in fiber length between the two interferometer arms.

The ranging performance of the FMCW LiDAR system using the FSL chirp source is shown in Fig.~\ref{figFMCW}b. The measured distance is plotted as a function of the reference distance for both first-order and second-order operation. In both cases, a clear linear relationship is observed, demonstrating accurate distance retrieval over the measured range.

\begin{table}[h]
\centering
\caption{Comparison of reference and measured distances for first- and second-order operation. Absolute and relative deviations are reported.}
\label{tab:ranging_results}
\begin{tabular}{c c c c c c}
\hline
\multirow{2}{*}{Reference distance} 
& \multicolumn{2}{c}{Measured distance } 
& \multicolumn{2}{c}{Deviation} \\
\cline{2-5}
& 1st order & 2nd order & 1st order & 2nd order \\
\hline
0~m & 0~m & 0~m & &  \\
0.50~m & 0.46~m & 0.50~m & -4~cm / 8~\% & 0~cm / 0 \% \\
0.70~m & 0.72~m & 0.71~m & 2~cm / 2.9 \% & 1~cm / 1.4 \% \\
0.80~m & 0.82~m & 0.79~m & 2~cm / 2.5 \% & -1~cm / 1.3 \% \\
0.96~m & 0.98~m & 0.94~m & 2~cm / 2.1 \% & -2~cm / 2.1 \% \\
1.00~m & 0.98~m & 1.00~m & -2~cm / 2.0 \% & 0~cm / 0 \% \\
1.04~m & 1.03~m & 1.04~m & -1~cm / 1.0 \% & 0~cm / 0 \% \\
2.00~m & 2.01~m & 1.97~m & 1~cm / 0.5 \% & -3~cm / 1.5 \% \\
3.00~m & 3.04~m & 2.98~m & 4~cm / 1.3 \% & -2~cm / 0.7 \% \\
\hline
\end{tabular}
\end{table}

Table~\ref{tab:ranging_results} summarizes the comparison between the reference distances and the distances measured using first- and second-order operation of the FSL-based FMCW LiDAR system. 
For both orders, good agreement with the reference values is observed over the full measurement range from 0 to 3~m. The absolute deviations remain within a few centimeters, corresponding to relative errors typically below 3 \%. 
The maximum achievable distance resolution is given by \(\Delta L = c/(2\Delta \nu)\), which yields approximately 5~cm for first-order operation with a chirp bandwidth of 3~GHz and approximately 1.5~cm for second-order operation with a bandwidth of 10~GHz. 
The experimentally observed deviations, which remain within a few centimeters over the full measurement range, are therefore in good agreement with the expected resolution limits. In particular, the reduced errors observed for second-order operation are consistent with its larger chirp bandwidth and correspondingly improved theoretical distance resolution.


These results confirm the high linearity and stability of the chirped waveforms generated by the frequency-shifted loop. 
The measured distances exhibit good agreement with the reference values over the investigated range, and the observed deviations are consistent with the theoretical resolution limits determined by the chirp bandwidth. 
Both first- and second-order operation enable accurate distance retrieval, with the second order benefiting from its larger bandwidth and improved resolution. 
The successful distance measurements demonstrate that the FSL concept provides a reliable and flexible platform for coherent ranging applications and validate its suitability for FMCW LiDAR systems.

\section{Conclusion}

We have experimentally demonstrated a recirculating frequency-shifting loop (FSL) architecture as a versatile platform for the generation of flexibly tunable optical chirps, which were employed for coherent distance measurements.

By using acousto-optic modulators (AOMs) and, in a second implementation, an electro-optic I/Q modulator operated in single-sideband (SSB) mode, we systematically investigated both low- and high-order frequency-shifting regimes and analyzed their impact on chirp dynamics.

For the AOM-based implementation, first- and second-order operation were characterized, revealing a strong and predictable dependence of the chirp rate on the modulation frequency, in good agreement with the analytical model. 
The limitations of our measurements are primarily of a practical nature. In our setup, the AOM drivers used allowed the output frequency around 130~MHz to be adjusted only in steps of 1~kHz, which restricted how closely the modulation frequency could be tuned to the exact mode-locking condition. With higher-resolution RF drivers, the detuning could be controlled more precisely, enabling operation significantly closer to the mode-locking condition. This would allow the generation of even steeper chirps, potentially reaching chirp rates that are difficult to achieve with any other technique.

Furthermore, time-frequency measurements confirmed the linear evolution of the instantaneous frequency within each pulse, and the extracted chirp rates followed the expected scaling behavior. In particular, the quadratic dependence on modulation frequency and the divergence near the mode-locking condition were experimentally verified.

Replacing the AOMs with an electro-optic I/Q modulator enabled access to higher-order operation up to the seventh order and increased achievable chirp rates, reaching several hundred megahertz per nanosecond. Although the usable chirp bandwidth was reduced due to the higher insertion loss of the I/Q modulator, this approach demonstrates the scalability of the FSL concept and its compatibility with integrated thin-film lithium niobate platforms.

Finally, the suitability of the FSL-generated chirps for coherent ranging was validated in an FMCW LiDAR experiment. The measured distances showed good agreement with reference values, and the observed deviations were consistent with the theoretical resolution limits determined by the chirp bandwidth. Both first- and second-order operation enabled accurate distance retrieval, with improved resolution achieved for larger chirp bandwidths.

Beyond coherent ranging, FSL implementations have demonstrated potential across diverse photonic domains. 
Full photonic integration of these configurations -- either via integrated acousto-optic modulators \cite{LoncarIntegratedAOM} or, more straightforwardly, via electro-optic modulators \cite{LoncarCMOS} in combination with integrated optical amplifiers \cite{KippenbergEDWA} -- could further unlock the potential of the deterministic linear frequency evolution intrinsically defined by the system's underlying mathematical framework. 

Overall, the FSL architecture provides a flexible and scalable platform for generating optical chirps with controllable bandwidth and chirp rate.
The demonstrated compatibility with electro-optic modulators opens a pathway toward integrated, high-bandwidth chirp sources for applications in coherent ranging, spectroscopy, microwave photonics, and high-speed optical signal processing.


\begin{backmatter}
\bmsection{Funding}
We acknowledge support by the Open Access Publication Fund of the University of Freiburg. This work was financially supported by the German Research Foundation, DFG (Grant No. BR 4194/12-1) and Federal Ministry of Research, Technology and Space, BMFTR (Grant No. 13N16555).

\bmsection{Acknowledgment}
We thank Prof. Dr. R\"udiger Quay (Fraunhofer IAF) for providing access to a 10~GHz oscilloscope.

\bmsection{Disclosures}
The authors declare no conflicts of interest.

\bmsection{Data Availability Statement}
The data that support the findings of this study are available from the corresponding author upon reasonable request.

\end{backmatter}


\bibliography{sample}

\end{document}